\documentclass[11pt,a4paper]{article}

\usepackage{amsmath,amssymb,ascmac,color,epsf,graphicx,latexsym,slashed,comment,fancyhdr,bm,cite,multirow,lscape}
\usepackage[colorlinks=true,linkcolor=blue,citecolor=blue,urlcolor=magenta]{hyperref} 

\newcommand{\figref}[1]{Figure~\ref{#1}}
\newcommand{\tblref}[1]{Table~\ref{#1}}
\newcommand{\secref}[1]{Section~\ref{#1}}

\newcommand{\appref}[1]{Appendix~\ref{#1}}
\newcommand{\ba}{\begin{align}}
\newcommand{\ea}{\end{align}}

\newcommand{\mc}[1]{\mathcal #1}
\newcommand{\mb}[1]{\mathbf #1}
\newcommand{\bs}[1]{\boldsymbol #1}

\newcommand{\nn}{\nonumber}
\newcommand{\mA}{\mathcal{A}}
\newcommand{\mB}{\mathcal{B}}
\newcommand{\Gcenter}[2]{
  \dimen0=\ht\strutbox
  \advance\dimen0\dp\strutbox
  \multiply\dimen0 by#1
  \divide\dimen0 by2
  \advance\dimen0 by-.5\normalbaselineskip
  \raisebox{-\dimen0}[0pt][0pt]{#2}}
\newcommand{\ma}[1]{({#1})$^2$}

\begin{document}
\begin{titlepage}
\begin{flushright}
UT--15--05\\
February, 2015
\end{flushright}

\vskip 4 cm
\begin{center}

{\fontsize{15.5pt}{0pt} \bf 
Diquark bound states with a completely crossed ladder truncation
}
\vskip 0.7in
{\large
\textbf{Ryusuke Jinno},
\textbf{Teppei Kitahara},
and
\textbf{Go Mishima}
}
\vskip 0.5in

{\large
{\it 
Department of Physics,  Faculty of Science, 
University of Tokyo, \\[0.4em]
Bunkyo-ku, 
Tokyo 113-0033, Japan
}}
\vskip 0.1in

\end{center}
\vskip .85in
\begin{abstract}
The Bethe-Salpeter equation in the diquark channel is investigated by employing the Dyson-Schwinger method 
together with the Munczek-Nemirovsky model.
The novelty of our study is a resummation of completely crossed ladder diagrams in the Bethe-Salpeter kernel.
These diagrams are enhanced due to their color factors in the diquark channel, but not in the meson channel.
In our analysis, diquark bound-state solutions exist in the Bethe-Salpeter equation.
\end{abstract}
\end{titlepage}
\renewcommand{\thefootnote}{\#\arabic{footnote}}
\setcounter{page}{1}
\hrule
\tableofcontents
\vskip .2in
\hrule
\vskip .4in

\section{Introduction}

One of the greatest challenges of particle physics is
to understand the dynamics and mass spectra of hadrons in terms of Quantum Chromodynamics (QCD),
and a concept of diquarks -- quark-quark bound states -- has been playing an important role towards these goals.
For example, diquark-quark models of baryon  \cite{GellMann:1964nj,Ida:1966ev,Lichtenberg:1967zz} are used to reproduce
the mass spectra, the decay rates, the magnetic moments, the ratio of the axial-vector coupling to the vector coupling,
the charge radius, and the processes of deep-inelastic scattering \cite{Anselmino:1992vg,Klempt:2009pi}.
Another example is phenomenology of recently confirmed exotic 
hadrons $X(3872),~Z_c(3900)^{\pm},~Z_c(4020)^{\pm},~Z(4430)^{\pm}$, etc.  \cite{Choi:2003ue,Ablikim:2013mio,Ablikim:2013emm,Mizuk:2009da,Aaij:2014jqa}.
It is proposed that a diquark-antidiquark description of such exotic hadrons can explain observed properties of those states, especially the narrow width and decay properties  
\cite{Weinberg:2013cfa,Knecht:2013yqa,Maiani:2014aja,Brodsky:2014xia,Esposito:2014rxa}. 

Meanwhile, a functional method using the Bethe-Salpeter (BS) equation with the Dyson-Schwinger (DS) equation 
has been considered to have no diquark bound-state solutions.
In the case of mesons, a consistent scheme which preserves the Nambu-Goldstone nature of pions is 
established for the extensively used rainbow-ladder (RL) truncation \cite{Maskawa:1974vs} and its extension  \cite{Bando:1993qy,Munczek:1994zz,Bender:1996bb},
and the resulting BS eq. has bound-state solutions in both truncations. 
However, 
in the diquark channel of the BS equation,
a certain kind of beyond-RL contributions have been found to remove the bound-state solutions \cite{Bender:1996bb,Hellstern:1997nv,Alkofer:2000wg,Bender:2002as}, in contrast to the RL truncation
\cite{Cahill:1987qr,Maris:2002yu}.

In this paper,
we investigate the differences among previous studies
\cite{Bender:1996bb,Hellstern:1997nv,Alkofer:2000wg,Bender:2002as} which make the diquark bound-state solutions absent,
and propose another approximation scheme in which diquark bound-state solutions exist.
To this end we employ the same QCD model as Refs.~\cite{Bender:1996bb,Alkofer:2000wg,Bender:2002as}, 
i.e., the Munczek-Nemirovsky model \cite{Munczek:1983dx}.
The simplicity of this model enables one to take higher loop diagrams into the BS kernel \cite{Bender:1996bb,Alkofer:2000wg,Bender:2002as,Maris:2003vk,Bhagwat:2004hn,Watson:2004kd,Matevosyan:2006bk,Gomez-Rocha:2014vsa,Matevosyan:2007cx,Matevosyan:2007wc},
leading the previous studies to find that a certain crossed ladder diagram gives a large repulsive contribution in the diquark BS equation due to its enhanced color factor \cite{Bender:1996bb,Alkofer:2000wg,Bender:2002as}.
We point out that this enhancement is not peculiar to the diagram considered in these studies,
but common among a class of diagrams -- the completely crossed ladder diagrams.
Then we perform a resummation of these diagrams adopting large-$N$ counting as a criteria for the truncation \cite{'tHooft:1974hx},
finding bound-state solutions to the diquark BS equation.
Although our calculation is based on a specific model with a specific truncation scheme, we believe that our result is 
at least a counterexample of the previous arguments that the BS eq. has no solutions in the diquark channel. 

The organization of this paper is as follows. 
In \secref{sec_DSE}, we introduce the DS eq. for quarks, and explain the model we adopt. 
Next in \secref{sec_BSE}, we introduce the BS eq. for mesons and diquarks, and discuss the type of diagrams which give 
the leading contribution to the BS kernel.
Our main analyses are presented in 
subsection \ref{sec_beyondG2} and \ref{sec_resum}.
The numerical solutions of the BS eq. and interpretations of them are shown in \secref{sec_discussion}, 
and the final section is devoted to conclusion.

\section{Dyson-Schwinger equation}
\label{sec_DSE}

Dyson-Schwinger (DS) equations are functional relations \cite{Dyson:1949ha,Schwinger:1951ex}
which are derived from a functional change of variables in the path integral in a modern point of view.
They include information on
non-perturbative effects in principle,
and provide us a useful way to study the infrared behavior of QCD.
The DS eq. of the quark propagator takes the form
\begin{align}
S^{-1}(p)
&= Z_2\left( {\rm i} \slashed{p} + m_b \right) +\Sigma (p), \label{eq_DSE1}\\
\Sigma (p)&= Z_1 g^2\int \! \frac{{\rm d}^4q}{(2\pi)^4} \ t^a \gamma_\mu S(q) t^a \Gamma_\nu (q,p)D_{\mu \nu}(k),
\label{eq_DSE2}
\end{align}
where $S(p)$ is the dressed quark propagator, $\Sigma (p)$ the quark self-energy, $D_{\mu\nu}$ the dressed gluon propagator,
$\Gamma_\mu$ the dressed quark-gluon vertex,   $m_b$ the bare mass of the current quark, 
$t^a$ the generator of SU(3)$_{\rm C}$, 
the gluon momentum $k = q - p$, 
$g$ is SU(3)$_{\rm C}$ gauge coupling constant, 
and  $Z_1$ and $Z_2$ are the renormalization factors of the quark-gluon vertex and the quark propagator, respectively. 
The general form of the dressed quark propagator is given as
\begin{align}
S(p)
&= -{\rm i}\slashed{p}\mc{A}(p^2) + \mc{B}(p^2).
\label{eq_AB}
\end{align}
The dependence on the renormalization scale $\zeta$ is omitted for simplicity.
Here we use the Euclidian metric and the gamma matrices satisfy $\{ \gamma_\mu, \gamma_\nu \} = 2\delta_{\mu \nu}$.
The dressed quark propagator is renormalized by the following condition:
\begin{align}
S^{-1}(p)|_{p^2 = \zeta^2}
&= {\rm i}\slashed{p} + m_0 (\zeta),
\end{align}
where $m_0(\zeta)$ is the current quark mass at the scale $\zeta$.

Equation \eqref{eq_DSE1} becomes a self-consistent equation of the quark propagator $S(p)$
when $\Gamma_\nu (p,q)$ and $D_{\mu\nu} (k)$ are modeled as some known functions of $p,q,k$ and $S(p)$.
This method can realize large quark mass at low energy, which is considered to be a consequence of 
the dynamical chiral symmetry breaking \cite{Maskawa:1974vs,Munczek:1983dx,Miransky:1983vj,Higashijima:1983gx}.

In this paper, we adopt the Munczek-Nemirovsky model  \cite{Munczek:1983dx},
\begin{align}
g^2 D_{\mu \nu} (k)
&= \left(  \delta_{\mu\nu}
- k_\mu k_\nu /k^2 \right) G (2\pi)^4 \delta^4(k)\nn\\
&= \genfrac{}{}{}{1}{3}{4} \delta_{\mu\nu}
G(2\pi)^4 \delta^4(k),
\label{eq_model}
\end{align}
where $G = \eta^2/4$ is a dimensionful parameter which determines the scale of this model.
In addition, the projection operator $(\delta_{\mu \nu} - k_\mu k_\nu / k^2)$ becomes $\frac{3}{4}\delta_{\mu\nu}$ due to the 
spherical symmetry of the model.
In this model, the three-point vertex of gluon vanishes because of its derivative coupling property, 
but the four-point vertex still remains.
Here and in what follows, we assume that the effects of the gluon self-coupling interactions
are not crucial for the calculation of the bound-state mass.
The validity of this assumption is not obvious
especially when we consider large-$N$ counting 
since vertex corrections due to these interactions are known to be
enhanced in the large-$N$ counting \cite{Matevosyan:2006bk,Bhagwat:2004kj,Alkofer:2008tt,Fischer:2009jm}.
We consider that the assumption is partially supported 
by the success for reproducing meson mass spectra in this setup
as shown in Refs.\cite{Munczek:1983dx,Bender:1996bb,Alkofer:2000wg,Bender:2002as,Maris:2003vk,Bhagwat:2004hn,Watson:2004kd,Matevosyan:2006bk,Gomez-Rocha:2014vsa}
and in Table \ref{tab_bound} of this paper.
We take the renormalization scale $\zeta$ to be large enough ($\zeta \gg 1 $ GeV), and
we can set the renormalization constants to $1$ 
and need not to specify the regularization scheme
in the practical calculation since the model has no UV divergence.
The dimensionful parameter $\eta$ can be fitted to reproduce the observed mass spectrum well,
and the value $\eta \simeq 1$ GeV is found to 
satisfy this requirement \cite{Munczek:1983dx}.
The $\delta$-function behavior in Eq.~\eqref{eq_model} is considered to 
be a realization of
the property of hadrons dominated by the infrared dynamics of quarks and gluons inside them.
Here the infrared region means the energy scale smaller than $G\simeq (0.5~\mathrm{GeV})^2$.

\begin{figure}
    \begin{center}
      \includegraphics[clip, width=9cm]{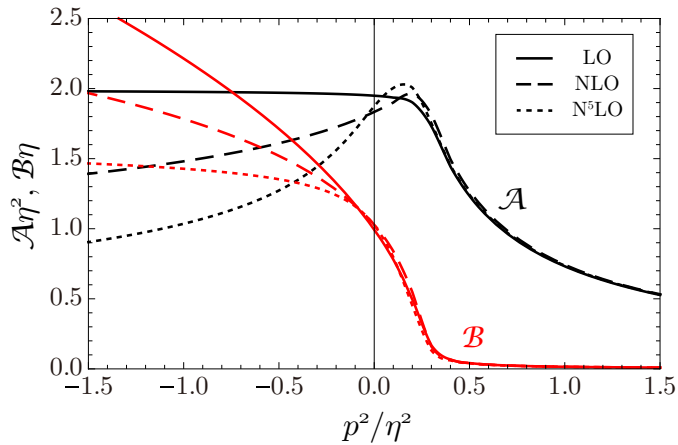}
    \end{center}  
     \vspace{-.2cm}
  \caption{\small 
  Solutions of the DS equation~\eqref{eq_DSE1}.
  The solid (dashed, dotted) line corresponds to the self-energy of
  \eqref{eq_SE_leading} (\eqref{eq_SE_NLO}, \eqref{eq_selfenergy} with $n=6$).
  The vertical axis is nondimensionalized by $\eta$.
  The solutions for $n=3,4,5$ in \eqref{eq_selfenergy} come between the red and yellow lines.
  See Appendix \ref{app_qDSE} for details.
  }
   \vspace{-.2cm}
\label{fig_DSE}
\end{figure}

In the Munczek-Nemirovsky model,
the self-consistency equation becomes simple algebraic equations
of ${\cal A}$ and ${\cal B}$, and thus can easily be solved once the self-energy $\Sigma (p)$ is specified.
One choice is the rainbow truncation,
\begin{align}
\Sigma(p)
&= G \gamma_\mu S(p) \gamma_\mu, 
\label{eq_SE_leading}
\end{align}
where the integration with respect to the loop momentum can easily be performed due to the $\delta$-function.
The quark propagator functions $\mA$ and $\mB$ are evaluated 
by solving Eq.~\eqref{eq_DSE1} with the self-energy Eq.~\eqref{eq_SE_leading},
and the results are plotted in \figref{fig_DSE}.
We take the current quark mass to be $m_0=0.012 \times \eta$ [GeV] in the plot,
which is consistent with previous studies.
With this choice, we obtain reliable masses of pion and rho mesons if we set $\eta \simeq 1$ GeV (see \tblref{tab_bound}).
More precise parameter fitting is possible but is beyond our scope; hence, we use this value for the current quark mass unless mentioned explicitly in the following calculations. 

\section{Bethe-Salpeter equation}
\label{sec_BSE}

\begin{figure}
    \begin{center}
      \includegraphics[clip, width=7cm]{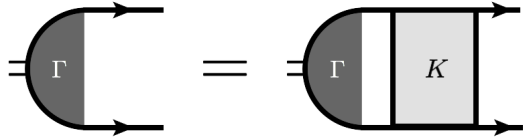}
    \end{center}
    \vspace{-.4cm}
  \caption{\small 
  Diagrammatic expression of the homogenous Bethe-Salpeter equation in a diquark channel.
  }
  \vspace{-.2cm}
\label{fig_bs}
\end{figure}

Bethe-Salpeter (BS) equation has been the central tool to investigate bound states
in terms of quantum field theory since its birth \cite{GellMann:1951rw,Salpeter:1951sz}.
It is a functional relation between four-point correlation functions,
and when we regard their residues as bound-state poles,
the BS eq. becomes a homogeneous equation with respect to a certain function -- BS amplitude.
The homogeneous BS eq. takes the following form:
\begin{align}
\Gamma (P,p)
&= \int \!\frac{{\rm d}^4k}{(2\pi)^4}\ K(P,p,k) S(k_+)\Gamma (P,k)S(k_-),
\label{eq_BS}
\end{align}
where $p$ and $P$ are the relative momentum and the total momentum of the system, respectively, 
$\Gamma (P,p)$ is the BS amplitude and $K(P,p,k)$, the BS kernel, is the $q\bar{q}$ scattering kernel 
which is two-particle-irreducible \cite{Cornwall:1974vz} with respect to $q\bar{q}$ lines, and
we define ${}^\forall l_\pm \equiv l \pm P/2$. 
The flavor, Dirac, and color indices are omitted in Eq.~\eqref{eq_BS} for simplicity. 
Note that the way to contract these indices depends on whether the bound state is meson or diquark.
\figref{fig_bs} is a diagrammatic expression of the homogenous Bethe-Salpeter equation in the diquark channel.

In the Munczek-Nemirovsky model,
the BS amplitude is a function of only $P$ because the momentum transfer by gluons is zero,
and its general form for the pseudoscalar meson channel is
\begin{align}
\Gamma (P) 
&= \gamma _5 \left[ f_1(P^2)+ {\rm i} \slashed{P} f_2(P^2) \right],
\label{eq_BSamp_sc}
\end{align}
while for the vector meson channel
\begin{align}
\Gamma _{\mu}^{\lambda} (P)
&= \epsilon^\lambda_\mu \left[ \gamma_\mu g_1(P^2)+ {\rm i} \sigma_{\mu \nu} P_\nu g_2(P^2) \right],
\label{eq_BSamp_vec}
\end{align}
where $\epsilon_{\mu}^{\lambda}~(\lambda=-1,0,+1)$ is the polarization four-vector and $\sigma _{\mu\nu}=[\gamma_\mu ,\gamma_\nu]/2$.
Substituting these into Eq.~\eqref{eq_BS}, one obtains a $2 \times 2$ matrix eigenvalue equation
\begin{align}
\left(
\begin{array}{c}
f_1\\ 
f_2
\end{array}
\right)
= H (P^2) 
\left(
\begin{array}{c}
f_1\\ 
f_2
\end{array}
\right),
\label{eq_eigeneq}
\end{align}
and a similar equation for the vector channel.
Note that there is no remaining integral due to the $\delta$-function in Eq.~\eqref{eq_model}. 
In order for Eq.~\eqref{eq_eigeneq} to have a nontrivial solution, the following condition:
\begin{align} 
\mathrm{det} \left[ \bm{I}-H(P^2) \right] = 0,
\label{eq_det}
\end{align}
with $\bm{I}$ being the $2 \times 2$ unit matrix must be satisfied.
Conversely, if Eq.~\eqref{eq_det} holds for a certain value of $P^2=-m^2<0$, then the BS eq. has a bound-state solution
and the pole mass of the state is given by $m$.
We treat Eq.~\eqref{eq_det} as an equation of one variable $P^2$,
and investigate whether it has a solution or not.

To evaluate the LHS of Eq.~\eqref{eq_det}, one has to specify the diagrams taken into the kernel $K$.
In the case of mesons, it is known that a consistent scheme for the DS and BS kernels exists which automatically preserves the axial-vector Ward-Takahashi identity
and keeps the Nambu-Goldstone nature of the resulting pion mass \cite{Maskawa:1974vs,Bando:1993qy,Bender:1996bb}.
In this construction, the BS kernel is determined to be
the sum of the terms with one of the quark propagators in the self-energy \eqref{eq_DSE2} replaced by
\begin{align}
S(p) 
&\to -S_+ \Gamma(P) S_-,
\label{eq_replace}
\end{align}
and with all the other quark propagators in the left (right) to the replaced one substituted by $S_+$ ($S_-$).
Here we defined $S_\pm = S(\pm P/2)$.
Following this scheme, one of the frequently adopted methods is to determine the self-energy \eqref{eq_DSE2} by the coupling expansion, 
and determine the diagrams in the BS kernel by Eq.~\eqref{eq_replace}.

On the other hand, no such symmetry is known for the diquarks. 
This means that the BS kernel does not necessarily have a correspondence to the DS kernel like Eq.~\eqref{eq_replace}.

In the subsection below, we examine the coupling expansion in both meson and diquark channels following previous studies.
This enables us to infer the kind of diagrams -- completely crossed diagrams -- 
which have the largest color factors at fixed orders of $G$.
Adopting a truncation scheme which takes these diagrams into account, 
we will see that the coupling expansion is unreliable in determining the existence of diquark bound-state solutions. 
Then we show that the resummation of all the completely crossed ladders, 
rather than the coupling expansion, leads to diquark bound-state solutions.

\subsection{Bound-state solutions up to ${\mc O}(G^2)$}
\label{sec_uptoG2}

\begin{figure}
    \begin{center}
      \includegraphics[clip, width=5cm]{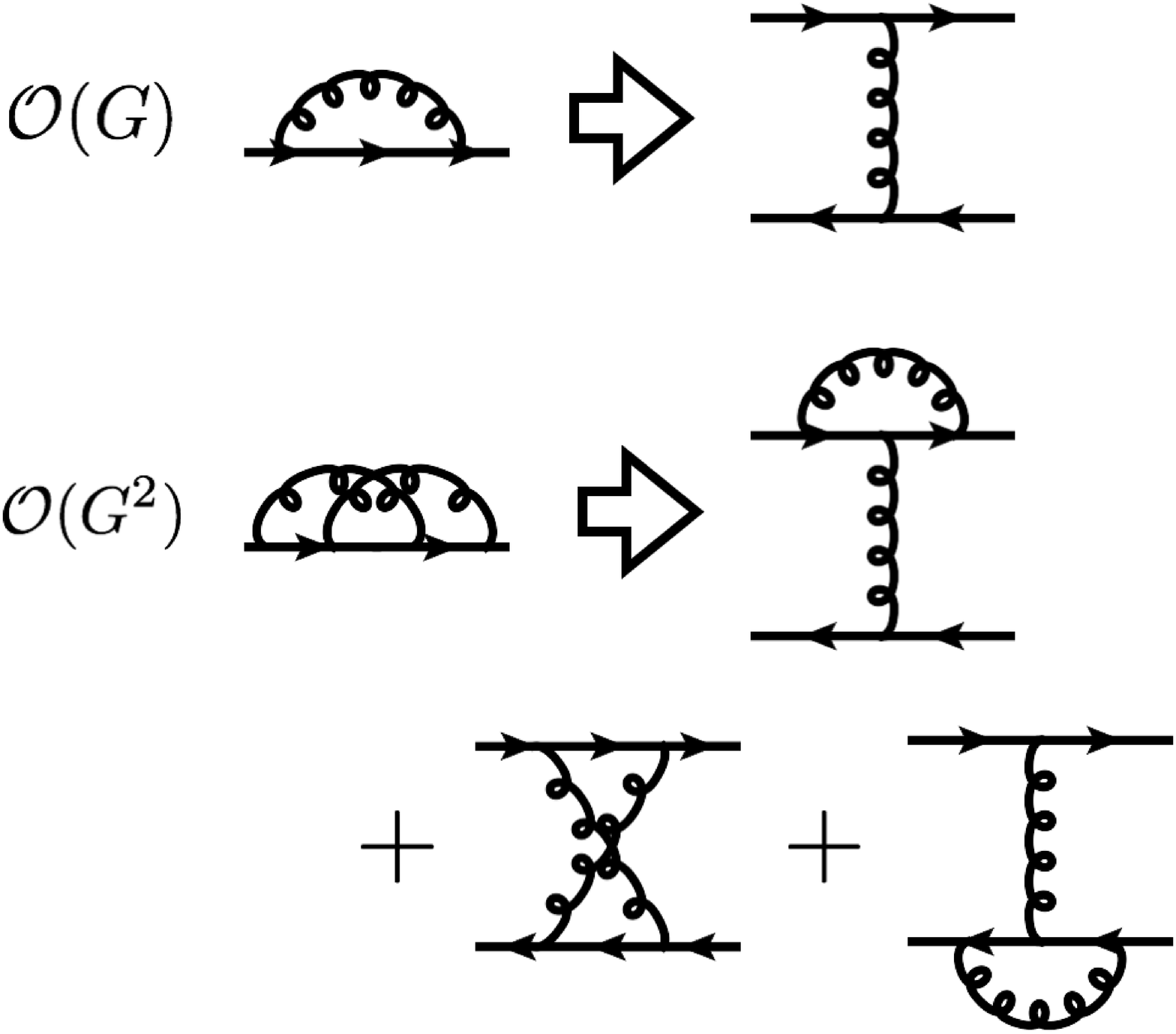}
    \end{center}  
    \vspace{-.6cm}
  \caption{\small 
  A diagrammatic expression of the Ward-Takahashi identity preserving combination of the quark Dyson-Schwinger eq.
  and the meson Bethe-Salpeter kernel in the rainbow-ladder approximation ($\mathcal{O}(G)$) and 
  next-to-leading order approximation ($\mathcal{O}(G^2)$).
  }
  \vspace{-.2cm}
\label{fig_consistent}
\end{figure}

We begin with ${\mc O}(G)$. 
The truncation of this order is called rainbow-ladder (RL) truncation.
Applying the prescription \eqref{eq_replace} to the quark self-energy shown in Eq.~\eqref{eq_DSE2}, 
we have the following BS eq. for flavor non-singlet mesons
\begin{align}
\Gamma (P) = 
-G\gamma _\mu S_+ \Gamma (P) S_- \gamma _\mu.
\label{eq_mBS_LO}
\end{align}
The BS kernel of Eq.~\eqref{eq_mBS_LO} is equal to the one obtained by the coupling expansion at this order.
The diquark BS eq. with the leading contribution to the BS kernel is
\begin{align}
\Gamma ^k (P) =-\genfrac{}{}{}{1}{1}{2} G\gamma _\mu S_+ \Gamma ^k (P) S_- \gamma _\mu.
\label{eq_dBS_RL}
\end{align}
Here some comments are in order for the diquark BS equation. 
The diquark BS amplitude has color indices due to its color non-singlet nature.
For example, the diquark BS eq. with this truncation becomes
\begin{align}
\Gamma _{ij} (P) = - \genfrac{}{}{}{1}{3}{4} G(t^a)_{ik} (t^{a})_{jl} \gamma _\mu S_+ \Gamma _{kl} (P) S_+^T \gamma _\mu ^T,
\label{eq_dBS1}
\end{align}
where $T$ denotes transposition, $i,j,k,l$ are color indices running from 1 to 3,
and $a$ is an adjoint index running from 1 to 8.
The Dirac matrices $S_+, \gamma_\mu$ and the color matrix $t^a$ are transposed 
in order to contract two quark lines correctly.
The scalar diquark BS amplitude is expressed as
\footnote{
Since color-symmetric ($\bm{6}$) diquarks do not form
bound states even in this truncation, 
we consider only color-antisymmetric ($\bar{\bm{3}}$) diquarks in this paper.}
\begin{align}
\Gamma _{ij} (P)
&= \epsilon _{ijk} \gamma _5 \left[ f_1^k (P^2)+ {\rm i} \slashed{P} f_2^k (P^2) \right] C
\equiv \epsilon _{ijk} \Gamma ^k (P) C,
\label{eq_BSamp_dq}
\end{align}
where $\epsilon_{ijk}$ is the antisymmetric tensor and 
$C=\gamma_2\gamma_4$ is the charge conjugation matrix.
It satisfies
\begin{align}
C^2=-1, \quad C\gamma_\mu ^TC=\gamma _\mu , \quad CS_+^TC=-S_-.
\label{eq_c}
\end{align}
Substituting Eq.~\eqref{eq_BSamp_dq} into Eq.~\eqref{eq_dBS1} and using Eqs.~\eqref{eq_diquark_pladderA} and \eqref{eq_c},
we obtain Eq.~\eqref{eq_dBS_RL}.

\begin{figure}
    \begin{center}
      \includegraphics[clip, width=7cm]{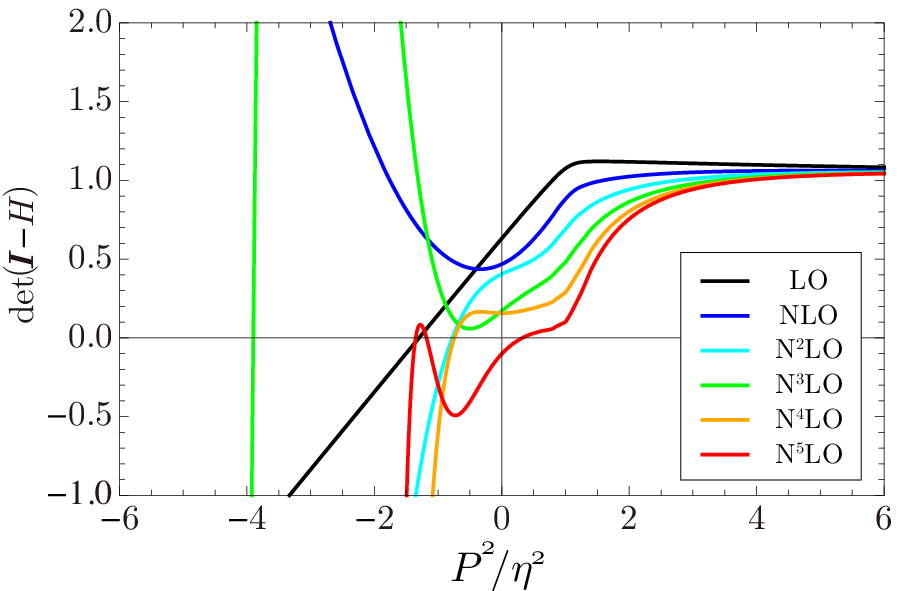}~~~~
      \includegraphics[clip, width=7cm]{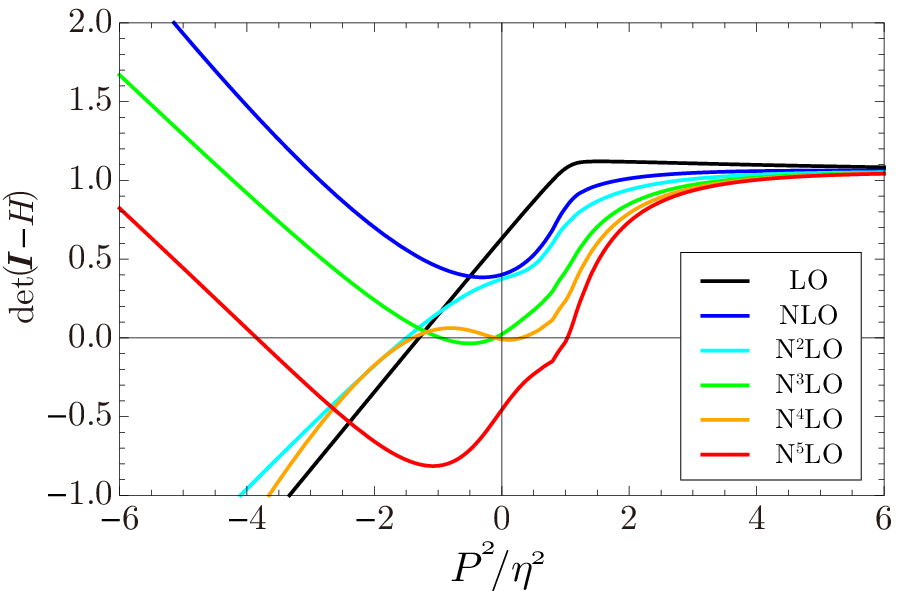}
    \end{center}  
    \vspace{-.2cm}
  \caption{\small
  The values of the det$(\bm{I}-H)$ in the diquark channel with Scheme A (left panel) and B (right panel).
  See the main text for the explanation of Scheme A and B in the diquark case.
  }
  \vspace{-.2cm}
\label{fig_diquarktot}
\end{figure}

We can set up the determinant equations for mesons and diquarks
following the procedure \eqref{eq_BSamp_sc}--\eqref{eq_det} 
using the quark DS eq. with the self-energy \eqref{eq_SE_leading}. 
At this order both BS eqs. have bound-sate solutions, and we summarize the resultant masses squared in the first column of \tblref{tab_bound}
as $\pi$, $\rho$ and $d(A)$.
In addition, we plot the behavior of the LHS of Eq.~\eqref{eq_det} for diquarks in \figref{fig_diquarktot}, 
which actually shows that the line has an intersection with the horizontal axis.
Note that the expression of the diquark BS eq. \eqref{eq_dBS_RL} has the same form as the meson one 
\eqref{eq_mBS_LO} except for the factor $\frac{1}{2}$. 
In general, the orders of $S_+, ~S_-, ~\gamma_\mu $ appearing in the meson and diquark BS eq. coincide at any order.

Next let us see ${\mc O}(G^2)$.
The quark self-energy is given by
\begin{align}
\Sigma =G\gamma_\mu S\gamma_\mu +\genfrac{}{}{}{1}{1}{8}G^2 \gamma_\mu S\gamma _\nu S\gamma _\mu S\gamma _\nu .
\label{eq_SE_NLO}
\end{align}
With the substitution \eqref{eq_replace}, the meson BS eq. becomes 
\begin{align}
\Gamma = 
&-G\gamma _\mu S_+ \Gamma S_- \gamma _\mu  \nn \\
&  -\genfrac{}{}{}{1}{1}{8}G^2 \gamma_\mu S_+ \left[ 
\Gamma S_-\gamma_\nu S_- \gamma _\mu   +\gamma_\nu S_+\Gamma S_-\gamma_\mu
+\gamma_\nu S_+\gamma_\mu S_+\Gamma 
\right]  S_-\gamma_\nu .
\label{eq_mBS_NLO}
\end{align}
Notice that there are three ${\mc O}(G^2)$ terms, since the replacement Eq.~\eqref{eq_replace} is applied to the three quark propagators
in Eq.~\eqref{eq_SE_NLO}.
The diagrammatic expression of the consistent combinations of the self-energy part of DS eq. and the meson BS kernel is shown in \figref{fig_consistent}.
Again, at this order, the BS kernel of Eq.~\eqref{eq_mBS_NLO} is found to be the same as 
the one obtained by the coupling expansion.
The diquark BS eq. at this order becomes
\begin{align}
\Gamma ^k 
=& -\genfrac{}{}{}{1}{1}{2}G\gamma _\mu S_+ \Gamma  ^kS_- \gamma _\mu \nn \\
 & -\genfrac{}{}{}{1}{1}{16}G^2 \gamma_\mu S_+ \left[ 
\Gamma ^kS_-\gamma_\nu S_- \gamma _\mu  +5\gamma_\nu S_+\Gamma ^kS_-\gamma_\mu 
+\gamma_\nu S_+\gamma_\mu S_+\Gamma ^k
\right]  S_-\gamma_\nu.
\label{eq_dBS_NLO}
\end{align}

\begin{table}
  \caption{\small 
  The values of bound-state mass squared, i.e., $-P^2$ in $\eta ^2$ [GeV$^2$] unit obtained by solving the Bethe-Salpeter equation 
  for pseudoscalar meson ($\pi$),
  vector meson ($\rho$), and
  scalar diquark ($d$) channels.
  Hyphen means that there is no bound-state solution
  and we list all the solutions with $-6<P^2<6$ 
  if the BS eq. has more than one solution.
  See the main text for the explanation of Scheme A, B in the diquark case.
  }
\label{tab_bound}
    \begin{center}
    \begin{tabular}{c||c|c|c|c|c|c}
    &$G$&$G^2$&$G^3$&$G^4$&$G^5$&$G^6$\\\hline\hline
    $\pi$ &\ma{0.136}&\ma{0.140}& \ma{0.142}& \ma{0.142}& \ma{0.142}& \ma{0.142}\\\hline
    $\rho$ &\ma{0.724}&\ma{0.793}& \ma{0.824}& \ma{0.849}& \ma{0.859}& \ma{0.842}\\\hline
    $d$ (A)&\ma{1.14}&-& \ma{0.88}&\ma{1.97}& \ma{0.86}&$-$\ma{0.58} \\
    &&&&&&\ \ \ma{1.09} \\
    &&&&&&\ \ \ma{1.16} \\\hline
    $d$ (B)&\ma{1.14}&-& \ma{1.22}&\ma{0.31} &$-$\ma{0.53}&$-$\ma{0.99}\\
    &&&&\ma{0.98}&\ \ \ma{0.34}&\ \ \ma{1.96} \\
    &&&&&\ \ \ma{1.18}&
    \end{tabular}
    \end{center}
\end{table}

Solving the determinant equation \eqref{eq_det} with the quark propagator calculated by Eq.~\eqref{eq_DSE1} using the self-energy 
Eq.~\eqref{eq_SE_NLO}, one finds that the bound-state solutions for both pseudoscalar and vector mesons still exist.
The resulting masses squared of these particles are listed in the second column of \tblref{tab_bound} as $\pi$ and $\rho$.
In both channels, the difference between the RL truncation \eqref{eq_mBS_LO} and the $\mc{O}(G^2)$ truncation \eqref{eq_mBS_NLO} is less than 10 \%, and one can see that these masses are rather stable against the coupling expansion.
In contrast, one finds that the bound-state solutions for diquarks disappear \cite{Bender:1996bb}. 
To see the disappearance of the solution, the value of the determinant \eqref{eq_det} calculated 
using Eq.~\eqref{eq_dBS_NLO} is plotted in \figref{fig_diquarktot}, where  
the plotted line has no intersection with the horizontal axis.

Here note that there is ambiguity about which truncation scheme in the quark self-energy should be used for diquarks, 
Eq~\eqref{eq_SE_leading} or Eq.~\eqref{eq_SE_NLO},
because unlike the meson case, no symmetry relates the quark DS eq. and the diquark BS kernel.
However, we confirmed that the disappearance of the diquark bound-state solution happens with both quark propagators; 
therefore, we do not discuss the quark DS eq. here.
In fact, as we show later, all the results below are qualitatively independent of the choice of the truncation used in the quark DS eq. 
as far as we use the quark propagators listed in this paper.

Where does the qualitative difference between mesons and diquarks come from?
Recall that the bound-state solution is present for mesons at this order.
The only difference in Eq.~\eqref{eq_mBS_NLO} and Eq.~\eqref{eq_dBS_NLO} is the factors on each term,
especially that of the second term of $\mc{O}(G^2)$.
This term comes from the crossed-ladder diagram, which can be considered to give a repulsive effect.
In order to check that, we solve the quark BS eq. taking only the crossed-ladder term in $\mc{O}(G^2)$ 
into account
\begin{align}
\Gamma ^k 
&= -\genfrac{}{}{}{1}{1}{2}G\gamma _\mu S_+ \Gamma  ^kS_- \gamma _\mu  -\genfrac{}{}{}{1}{5}{16}G^2 \gamma_\mu S_+ 
\gamma_\nu S_+\Gamma ^kS_-\gamma_\mu 
 S_-\gamma_\nu ,
\label{eq_dBS_NLO2}
\end{align}
and obtain qualitatively the same result. 

Note that the enhancement of the crossed ladder diagram is not peculiar to ${\mc O}(G^2)$.
To see this, let us see the color factors of (crossed) ladder diagrams at $\mc{O}(G^n)$.
The ladder diagram for the meson channel has a factor (see \appref{app_colfac})
\begin{align}
t^{a_1}t^{a_2}\cdots t^{a_{n-1}}t^{a_n}t^{a_n}t^{a_{n-1}}\cdots t^{a_2}t^{a_1} \sim N^n/2^n,
\label{eq_meson_ladder}
\end{align}
whereas the completely crossed ladder diagram has
\begin{align}
t^{a_1}t^{a_2}\cdots t^{a_{n-1}} t^{a_n}t^{a_1}t^{a_2}\cdots t^{a_{n-1}} t^{a_n}\sim N/2^n,
\label{eq_meson_pladder}
\end{align}
in the large-$N$ limit.
The enhancement of the ladder diagrams and the relative suppression of others including the completely crossed ladder diagrams 
provide one ground for the RL approximation.
On the other hand, the ladder diagram for the diquark channel has
\begin{align}
t^{a_1}t^{a_2}\cdots t^{a_{n-1}}t^{a_n}\epsilon t^{Ta_n} t^{Ta_{n-1}}\cdots t^{Ta_2}t^{Ta_1}
\sim \epsilon /2^n,
\label{eq_diquark_ladder}
\end{align}
whereas the completely crossed ladder diagram has
\begin{align}
t^{a_1}t^{a_2}\cdots t^{a_{n-1}} t^{a_n}\epsilon t^{Ta_1}t^{Ta_2}\cdots t^{Ta_{n-1}}t^{Ta_n}
\sim \epsilon N^{n-1}/2^n,
\label{eq_diquark_pladder}
\end{align}
where $\epsilon$ is a symbolic expression of the antisymmetric tensor $\epsilon _{ijk}$. 

What we find from these calculations is that the crossed-ladder contributions 
are not perturbations in the case of the diquark BS equation.
Although the argument based on the large-$N$ limit may not be directly applicable to the real-world QCD with $N=3$,  
the enhancement found above indicates the possibility that an inclusion of higher order terms of this type changes the result again.

\subsection{Bound-state solutions beyond ${\mc O}(G^2)$}
\label{sec_beyondG2}
\begin{figure}
    \begin{center}
      \includegraphics[clip, width=4cm]{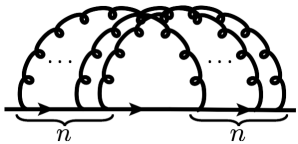}
    \end{center}  
    \vspace{-.4cm}
  \caption{\small 
  Diagrammatic expression of the quark self-energy of
  the completely crossed gluon correction \eqref{eq_selfenergy}.
    }
    \vspace{-.2cm}
\label{fig_selfenergy}
\end{figure}

First we investigate the effect of the completely crossed ladder diagrams on the meson BS eq. beyond ${\mc O}(G^2)$ for comparison with that on the diquark BS equation. 
In order to realize both the inclusion of the completely crossed ladders and the symmetry preserving scheme explained in Eq.~\eqref{eq_replace},
we choose the self-energy of the quark DS eq. to be 
\begin{align}
\Sigma &=-\sum _{n} (-G)^n  \left(\genfrac{}{}{}{1}{3}{4}\right)^n d_n^{({\rm M})} \gamma _{\mu_1}S\gamma _{\mu_2}S\cdots S\gamma _{\mu_n}S\gamma _{\mu_1}S\cdots S\gamma _{\mu_n},
\label{eq_selfenergy}
\end{align}
with 
\begin{align}
d_n^{({\rm M})}&= \frac{N-\genfrac{}{}{}{1}{1}{N}}{2^{n+1}} \left[ \left(-1-\genfrac{}{}{}{1}{1}{N} \right) ^{n-1}+\left( 1-\genfrac{}{}{}{1}{1}{N} \right) ^{n-1}\right],
\end{align}
and \figref{fig_selfenergy}~is a diagrammatic expression of it.
Then the replacement \eqref{eq_replace} determines the BS kernel.
At ${\mc O}(G^n)$,
there are $2n-1$ corresponding terms in the BS equation,
and we can solve the BS eq. in exactly the same way as in the previous subsection. 
We put the actual calculation in Appendix \ref{app_qDSE}.
The results up to ${\mc O}(G^6)$ are summarized in \tblref{tab_bound},
which shows the effect of the completely crossed ladders is insignificant in the meson channel.

Next we investigate the diquark BS eq. with the completely crossed ladder diagrams.
It has a form
\begin{align}
\Gamma _k
&=- \sum _n (-G)^n \left( \genfrac{}{}{}{1}{3}{4} \right) ^n d^{({\rm D})} _n \gamma _{\mu_1}S_+ \cdots S_+ \gamma _{\mu_n}S_+ \Gamma _k S_- \gamma _{\mu_1}S_- \cdots S_- \gamma _{\mu_n}, \label{eq_BSEdiquark}
\end{align}
with
\begin{align}
d^{({\rm D})} _n&= \genfrac{}{}{}{1}{1}{2^n}  \left[ (-1)^n (N+1) \genfrac{}{}{}{1}{1}{N^{n+1}} -\genfrac{}{}{}{1}{1}{N} \left( N-\genfrac{}{}{}{1}{1}{N} \right) ^{n} \right].
\end{align}
There are some options for the quark self-energy similar to what we have done in the $\mc{O} (G^2)$ case.
In order to investigate the truncation dependence, we try two schemes, Eqs.~\eqref{eq_SE_leading} and \eqref{eq_selfenergy}, for the quark self-energy, and call them Scheme A and B, respectively.
Scheme A is chosen to include the leading-order contribution in the 't~Hooft limit \cite{'tHooft:1974hx}, 
i.e. the large-$N$ limit with $G=\tilde G/N$,
which is consistent with the criteria used for the resummation of completely crossed ladder diagrams in the next subsection.
Scheme B is the same one for mesons at each order of $G$.
The determinant \eqref{eq_det} for these diquark BS eqs. are plotted in \figref{fig_diquarktot} up to ${\mc O}(G^6)$,
and the bound-state masses squared (if the solution exists) are summarized in \tblref{tab_bound}.
When the BS eq. has more than one solution with $-6<P^2<6$, we list all of them in \tblref{tab_bound}.
As discussed in the previous subsection,
the bound-state solution present in the ${\mc O}(G)$ truncation (the RL truncation)
disappears at ${\mc O}(G^2)$.
However, it revives at ${\mc O}(G^3)$
and somehow persists up to ${\mc O}(G^6)$.
In \figref{fig_diquarktot}, one finds that the behavior of the determinant \eqref{eq_det} at time-like region changes qualitatively 
as the order of $G$ increases, especially that the sign of the gradient changes alternately. 
We try taking account of other diagrams in the diquark BS kernel by imitating the meson BS kernel constructed below Eq.~\eqref{eq_selfenergy},
and obtain qualitatively the same result.

From the enhancement of the color factors discussed in Eqs.~\eqref{eq_meson_ladder}--\eqref{eq_diquark_pladder} and the results displayed above, 
one possible origin of instability against the coupling expansion in the diquark BS eq. is the completely crossed ladder diagrams, 
which have the largest color factors at fixed orders of $G$.
In the following, we provide a way to take all these diagrams into account.

\subsection{Resummation of the completely crossed ladder diagrams}
\label{sec_resum}

As we see in the previous subsection, the coupling expansion seems not to be valid
in the diquark BS kernel due to their enhanced color factors in the completely crossed ladder diagrams.
Since the completely crossed ladders are found not to be perturbations, 
we perform a resummation of these diagrams in the diquark BS kernel.
Note that all of these diagrams are the leading contribution in light of 
the 't~Hooft limit \cite{'tHooft:1974hx}, namely, the large-$N$ limit with $G=\tilde G/N$.
Therefore, our resummation can be regarded as a truncation with respect to the 't~Hooft limit.

The basis for the BS kernel is spanned by 16 elements, which we choose as follows:
\begin{align}
B = \{ b_I | I = 1,2,...,16 \}, 
\end{align}
where
\begin{align}
\begin{array}{llll}
b_1 
= \bs{1} \otimes \bs{1} , &
b_5 
= \gamma_\mu \otimes \gamma_\mu , &
b_9 
= \gamma_\mu\gamma_\nu \otimes \gamma_\mu\gamma_\nu , &
b_{13}
= \gamma_\mu\gamma_\nu\gamma_\rho \otimes \gamma_\mu\gamma_\nu\gamma_\rho , \\
b_2
= \bs{1} \otimes \slashed{P} , &
b_6 
= \gamma_\mu \otimes \slashed{P}\gamma_\mu , &
b_{10} 
= \gamma_\mu\gamma_\nu \otimes \slashed{P}\gamma_\mu\gamma_\nu , &
b_{14}
= \gamma_\mu\gamma_\nu\gamma_\rho \otimes \slashed{P}\gamma_\mu\gamma_\nu\gamma_\rho , \\
b_3 
= \slashed{P} \otimes \bs{1} , &
b_7 
= \slashed{P}\gamma_\mu \otimes \gamma_\mu , &
b_{11}
= \slashed{P}\gamma_\mu\gamma_\nu \otimes \gamma_\mu\gamma_\nu , &
b_{15}
= \slashed{P}\gamma_\mu\gamma_\nu\gamma_\rho \otimes \gamma_\mu\gamma_\nu\gamma_\rho , \\
b_4 
= \slashed{P} \otimes \slashed{P} , &
b_8 
= \slashed{P}\gamma_\mu \otimes \slashed{P}\gamma_\mu ,&
b_{12}
= \slashed{P}\gamma_\mu\gamma_\nu \otimes \slashed{P}\gamma_\mu\gamma_\nu , &
b_{16}
= \slashed{P}\gamma_\mu\gamma_\nu\gamma_\rho \otimes \slashed{P}\gamma_\mu\gamma_\nu\gamma_\rho.\\
\end{array}
\label{eq_basis}
\end{align}
Other elements than those shown here (e.g. $\gamma_\mu \gamma_\nu \gamma_\rho \gamma_\sigma \otimes \gamma_\mu \gamma_\nu \gamma_\rho \gamma_\sigma$) can be expressed by the ones shown here, as proven in \appref{app_proof}.
Here $b_I = u_I \otimes l_I $ is understood as $(b_I)_{\alpha_1\alpha_2,\beta_1\beta_2} = (u_I)_{\alpha_1\alpha_2}(l_I )_{\beta_1\beta_2}$, where $\alpha_1,~\alpha_2,~\beta_1,~\beta_2$ are Dirac indices.
Note that the Lorentz indices are contracted between the upper ($u_I$) and lower ($l_I$) halves of the basis, 
since both are connected by gluon lines, which are effectively proportional to $\delta_{\mu \nu}$.
Also note that the only momentum appearing above is $P$, the total momentum of the diquark, 
since the gluon momentum vanishes in the present model.

\begin{figure}
    \begin{center}
      \includegraphics[clip, width=7cm]{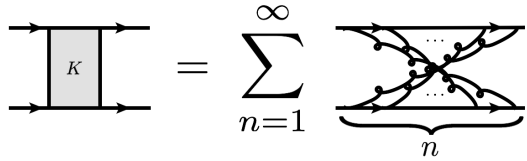}
    \end{center}
    \vspace{-.8cm}
  \caption{\small 
  Diagrammatic expression of the resummation of the completely crossed ladder diagrams performed in subsection \ref{sec_resum}.
  }
  \vspace{-.2cm}
\label{fig_kernelrec}
\end{figure}

We define $K_n$ as the completely crossed diagram in the diquark channel, 
with $n$ gluon lines propagating between two quark propagators. 
Then our choice of BS kernel is $K=\sum_{n=1}^\infty K_n$ (see \figref{fig_kernelrec}). 
Let us factorize $K_n$ as $K_n = (-G)^n(\frac{3}{4})^n d^{({\rm D})}_n \tilde{K}_n$, 
where $d^{({\rm D})}_n$ is the color factor defined in \appref{app_colfac}.
The vertex and projection factor $(-G)^n (\frac{3}{4})^n$, and the color factor $d^{({\rm D})}_n$ are taken into account at the end of the calculation.
We can relate $\tilde{K}_{n+1}$ to $\tilde{K}_n$ by one-gluon crossing,
the effect of which can be expressed by an operator $\mc{R}$:
\begin{align}
\tilde{K}_{n+1} 
&= \mc{R}[\tilde{K}_n].
\end{align}
Now let us expand $\tilde{K}_n$ as
\begin{align}
\tilde{K}_n
&= \bs{\alpha}_n \cdot \bs{b}
= \sum_{I=1}^{16} \alpha_{n,I} b_I,
\end{align}
where $\alpha_{n,I}$ are coefficients of each basis $b_I$.
For example, $\tilde{K}_1 = -\gamma_\mu \otimes \gamma_\mu$, so that $a_{1,I} = -\delta_{5I}$.
The operation of $\mc{R}$ on each basis $b_I$ is expressed as
\begin{align}
\mc{R}[b_I]
&= -\left(  \gamma_\mu S_+ u_I \otimes \gamma_\mu S_- l_I \right),
\end{align}
where the factor $(-1)$ comes from the charge conjugation explained in Eq.~\eqref{eq_c}.
Defining the recursion matrix $R$ by\footnote{
In the practical calculation, we define an inner product $I$ among $b_I$'s as 
\begin{align}
I(b_I,b_J)
&= {\rm Tr}_D[u_I u_J s l_I l_J s] +  {\rm Tr}_D[u_I u_J s l_J l_I s]
\end{align}
with $s=1+2\gamma^5$, to invert Eq.~(\ref{eq_basisrec}) to obtain $R_{IJ}$. 
Note that the choice of $s$ is arbitrary as long as the rank of the matrix $I(b_I,b_J)$  is 16.
}
\begin{align}
R_{JI}b_J
&= \mc{R}[b_I],
\label{eq_basisrec}
\end{align}
the recursion relation between $K_n$ and $K_{n+1}$
gives
\begin{align}
\alpha_{n+1,I}
&= R_{IJ}\alpha_{n,J}.
\end{align}
Note that the matrix $R$ is a function of $\mA(s/4),\mB(s/4)$ and $s=P^2$. 
The explicit form of the matrix $R_{IJ}$ is shown in \tblref{tbl_R}.
The coefficient vector $\mb{\alpha}_n$ becomes
\begin{align}
\bs{\alpha}_n
&= R^{n-1}\bs{\alpha}_1,
\end{align}
and thus the sum $K$ of the completely crossed diagrams is given by
\begin{align}
K = \sum_{n=1}^\infty K_n  
&= \sum_{n=1}^\infty (-G)^n \left( \genfrac{}{}{}{1}{3}{4} \right)^n d^{({\rm D})}_n \tilde{K}_n \nn\\
&= \sum_{n=1}^\infty (-G)^n \left( \genfrac{}{}{}{1}{3}{4} \right)^n d^{({\rm D})}_n \bs{\alpha}_n \cdot \bs{b}\nn\\
&= G\left[ \left( \left[ (-G)\left( \genfrac{}{}{}{1}{3}{4} \right) R+6 \right]^{-1} - \left[ (-G)\left( \genfrac{}{}{}{1}{3}{4} \right) 4R-3\right]^{-1} \right)  \bs{\alpha}_1 \right] \cdot \bs{b},
\end{align}
where we used the color factor shown in Eq.~(\ref{eq_diquark_pladderA}).

\begin{table}[tbp]
\caption{\small Recursion matrix $R_{IJ}$.}
\label{tbl_R}
\begin{center}
\scalebox{0.9}[0.9]{
\begin{tabular}{c|c||cccccccc}
\multicolumn{2}{c||}{\Gcenter{2}{$R_{IJ}$}} & \multicolumn{8}{c}{$J$} \\\cline{3-10}
\multicolumn{2}{c||}{} & 1 & 2 & 3 & 4 & 5 & 6 & 7 & 8 \\\hline \hline
\multirow{16}{*}{$I$}
& \multirow{1}{*}{1} &&&&&&&\\
& \multirow{1}{*}{2} &$i\mA\mB$&$-\mA^2s/2$&$-2\mB^2$&$-i\mA\mB s$&&&&\\
& \multirow{1}{*}{3} &$-i\mA\mB$&$-2\mB^2$&$-\mA^2s/2$&$i\mA\mB s$&&&&\\
& \multirow{1}{*}{4} &&&&&&&\\
& \multirow{1}{*}{5} &$-\mB^2$&$-i\mA\mB s/2$&$i\mA\mB s/2$&$-\mA^2s^2/4$&&&&\\
& \multirow{1}{*}{6} &$i\mA\mB/2$&$\mB^2$&$\mA^2 s/4$&$-i\mA\mB s/2$&$i\mA\mB$&$-\mA^2 s/2$&$-2\mB^2$&$-i\mA\mB s$\\
& \multirow{1}{*}{7} &$-i\mA\mB/2$&$\mA^2 s/4$&$\mB^2$&$i\mA\mB s/2$&$-i\mA\mB$&$-2\mB^2$&$-\mA^2 s/2$&$i\mA\mB s$\\
& \multirow{1}{*}{8} &$-\mA^2/4$&$i\mA\mB/2$&$-i\mA\mB/2$&$-\mB^2$&&&&\\
& \multirow{1}{*}{9} &&&&&$-\mB^2$&$-i\mA\mB s/2$&$i\mA\mB s/2$&$-\mA^2 s^2/4$\\
& \multirow{1}{*}{10} &&&&&$i\mA\mB/2$&$\mB^2$&$\mA^2 s/4$&$-i\mA\mB s/2$\\
& \multirow{1}{*}{11} &&&&&$-i\mA\mB/2$&$\mA^2 s/4$&$\mB^2$&$i\mA\mB s/2$\\
& \multirow{1}{*}{12} &&&&&$-\mA^2/4$&$i\mA\mB/2$&$-i\mA\mB/2$&$-\mB^2$\\
& \multirow{1}{*}{13} &&&&&&&&\\
& \multirow{1}{*}{14} &&&&&&&&\\
& \multirow{1}{*}{15} &&&&&&&&\\
& \multirow{1}{*}{16} &&&&&&&&\\
\multicolumn{2}{c}{} &&&&&&&&\\
\multicolumn{2}{c||}{\Gcenter{2}{$R_{IJ}$}} & \multicolumn{8}{c}{$J$} \\\cline{3-10}
\multicolumn{2}{c||}{} & 9 & 10 & 11 & 12 & 13 & 14 & 15 & 16 \\\hline \hline
\multirow{16}{*}{$I$}
& \multirow{1}{*}{1} &&&&&&&\\
& \multirow{1}{*}{2} &&&&&&&&\\
& \multirow{1}{*}{3} &&&&&&&&\\
& \multirow{1}{*}{4} &&&&&&&\\
& \multirow{1}{*}{5} &&&&&$4\mA^2 s$&$-8i\mA\mB s$&$8i\mA\mB s$&$16\mB^2 s$\\
& \multirow{1}{*}{6} &&&&&$8i\mA\mB$&$-4\mA^2 s$&$-16\mB^2$&$-8i\mA\mB s$\\
& \multirow{1}{*}{7} &&&&&$-8i\mA\mB$&$-16\mB^2$&$-4\mA^2 s$&$8i\mA\mB s$\\
& \multirow{1}{*}{8} &&&&&$16\mB^2/s$&$8i\mA\mB$&$-8i\mA\mB$&$4\mA^2 s$\\
& \multirow{1}{*}{9} &$-4\mB^2$&$-2i\mA\mB s$&$2i\mA\mB s$&$-\mA^2 s^2$\\
& \multirow{1}{*}{10} &$i\mA\mB$&$-\mA^2 s/2$&$-2\mB^2$&$-i\mA\mB s$&$2i\mA\mB$&$4\mB^2$&$\mA^2 s$&$-2i\mA\mB s$\\
& \multirow{1}{*}{11} &$-i\mA\mB$&$-2\mB^2$&$-\mA^2 s/2$&$i\mA\mB s$&$-2i\mA\mB$&$\mA^2 s$&$4\mB^2$&$2i\mA\mB s$\\
& \multirow{1}{*}{12} &&&&&$-\mA^2$&$2i\mA\mB$&$-2i\mA\mB$&$-4\mB^2$\\
& \multirow{1}{*}{13} &$-\mB^2$&$-i\mA\mB s/2$&$i\mA\mB s/2$&$-\mA^2 s^2/4$&$-\mA^2s$&$2i\mA\mB s$&$-2i\mA\mB s$&$-4\mB^2 s$\\
& \multirow{1}{*}{14} &$i\mA\mB/2$&$\mB^2$&$\mA^2s/4$&$-i\mA\mB s/2$&$-i\mA\mB$&$\mA^2s/2$&$2\mB^2$&$i\mA\mB s$\\
& \multirow{1}{*}{15} &$-i\mA\mB/2$&$\mA^2s/4$&$\mB^2$&$i\mA\mB s/2$&$i\mA\mB$&$2\mB^2$&$\mA^2 s/2$&$-i\mA\mB s$\\
& \multirow{1}{*}{16} &$-\mA^2/4$&$i\mA\mB/2$&$-i\mA\mB/2$&$-\mB^2$&$-4\mB^2/s$&$-2i\mA\mB$&$2i\mA\mB$&$-\mA^2s$
\end{tabular}}
\end{center}
\end{table}

\section{Numerical results and discussion}
\label{sec_discussion}

\begin{figure}
    \begin{center}
      \includegraphics[clip, width=9cm]{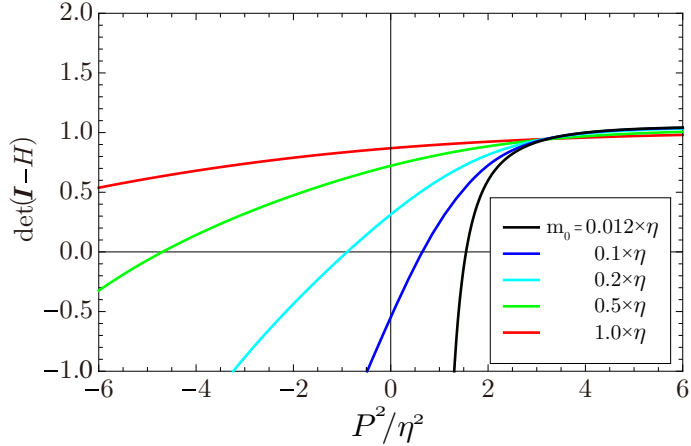}
    \end{center}
    \vspace{-.2cm}
  \caption{\small 
  Values of the determinant Eq.~\eqref{eq_det} obtained by the resummation of the completely crossed ladder diagrams
  performed in subsection \ref{sec_resum} for various current quark masses.
  Each line corresponds to the current quark mass $m_0 = 0.012,~0.1,~0.2,~0.5,~1.0 \times \eta$~[GeV].
  }
  \vspace{-.2cm}
\label{fig_det}
\end{figure}

After the resummation of the completely crossed ladder diagrams,
the diquark BS eq. actually has a bound-state solution.
Figure~\ref{fig_det} is a plot of det$(\bm{I}-H(P^2))$ in the scalar diquark channel as a function of $P^2$ for various current quark masses $m_0$.
The figure shows that the BS eq. does have solutions of det$(\bm{I}-H)=0$. 
This is the main result of our study.

\begin{figure}
    \begin{center}
      \includegraphics[clip, width=9cm]{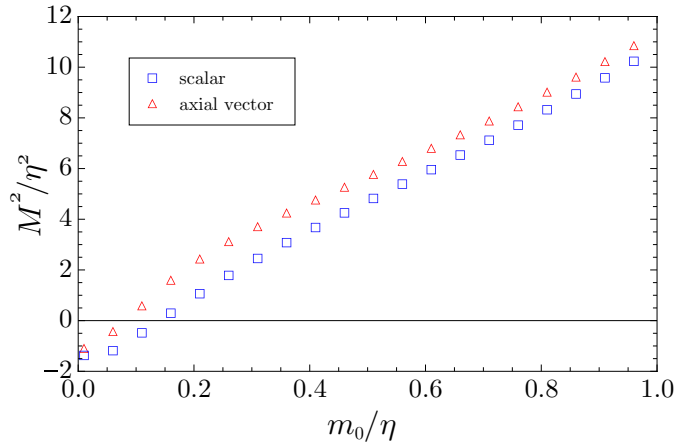}
    \end{center}
    \vspace{-.4cm}
  \caption{\small 
  Diquark masses as a function of the input current quark mass.
  These values are obtained by the resummation of the completely crossed ladder diagrams
  performed in subsection \ref{sec_resum}.
  }
  \vspace{-.2cm}
\label{fig_diquarkmass}
\end{figure}

We observe similar behavior in the axial-vector diquark channel, and the resulting masses are summarized in \figref{fig_diquarkmass}.
The axial vector diquarks tend to be heavier than the scalar diquarks, which is consistent with other studies 
\cite{Cahill:1987qr,Maris:2002yu,Hess:1998sd,Alexandrou:2006cq,Kleiv:2013dta}.

There are two subtleties in our calculation that have to be discussed.
One is the ambiguity of the truncation scheme for the quark self-energy,
as we have mentioned repeatedly in the previous section.
We check that the existence of bound-state solutions is unchanged under different truncation schemes for the DS equation, 
although the values of bound-state masses change.
The results in \figref{fig_det} and \figref{fig_diquarkmass} are calculated with the quark self-energy
Eq.~\eqref{eq_SE_leading}, since it is the leading term in the 't~Hooft limit.
The other is the space-like poles appearing when the current quark mass is $m_0 \lesssim 0.1\times \eta$ [GeV].
This can be a consequence of the simplification of the QCD model and the truncation scheme we adopt,
and there is a possibility that other models or truncation schemes resolve this artifact.

We would like to note that the presence of the diquark bound state does not contradict its confined nature.
In the previous studies \cite{Bender:1996bb,Alkofer:2000wg,Bender:2002as} 
the absence of diquark bound states is regarded as a realization of confinement.
This argument is based on a fact that correlation functions without one-particle poles do not contribute to the S-matrix.
However, there is another realization of confinement 
based on the BRST quartet mechanism \cite{Kugo:1979gm}, 
in which it is proven that under a certain condition color non-singlet states are also BRST non-singlet states, and hence they do not appear in the physical Hilbert space.
Therefore, if quarks are confined by this mechanism, then diquarks are confined simultaneously.

Our result that the diquark BS eq. has a bound-state solution gives a natural explanation 
to phenomenologically successful diquark constituent models:
first two of three (or four) quarks form a bound state and can be treated as a one-particle state.
However, we should note that the absence of bound-state solutions to the diquark BS eq. itself does not 
contradict to these models.
What we have investigated in this paper (and in other papers about the diquark BS eq.) 
is merely a pure two quark system, after all.
There is a possibility that more complex dynamics happening inside hadrons makes
two quarks bound even if their pure BS eq. has no bound-state solutions.

We also comment on our interpretation of the completely crossed ladders.
In weakly bound systems like the hydrogen atom, the constituent particles (proton and electron)
can be regarded as nearly on-shell, and
the leading effect of the exchange of photon becomes instantaneous interaction.
In this case, the completely crossed diagrams of higher orders require the constituent particles
to become off-shell by the energy-momentum conservation,
which leads to suppression of these effects \cite{Berestetsky:1982aq}.
However, in the case of diquarks and mesons, a naive application of above discussion is impossible
because the on-shellness of quarks inside them are no longer meaningful.
This interpretation is valid if the current masses of quarks are small.
The reason why the coupling expansion of the meson BS kernel is successful
may be the suppression of the color factor discussed around Eqs.~\eqref{eq_meson_ladder}-\eqref{eq_diquark_pladder}.
To summarize, we suppose that the completely crossed ladders can be neglected
when the constituents are regarded as nearly on-shell (hydrogen atom) or
when their coefficients are suppressed (meson),
and diquark is an exception of both cases.

\section{Conclusion}
\label{sec_conc}

The diquark Bethe-Salpeter equation with the Munczek-Nemirovsky model is investigated.
We find that the contributions of completely crossed ladder diagrams in the diquark channel 
affect the behavior of the Bethe-Salpeter kernel significantly, in contrast to the meson channel. 
We perform a resummation of all the completely crossed ladder diagrams in the Bethe-Salpeter kernel. 
As a result, we find that diquark bound-state solutions exist.
Although the values of resulting diquark masses may contain model- and truncation-artifacts 
and are far from a reliable prediction at this stage,
the presence of the bound-state solution is rather stable against variations in the parameter values and truncation schemes.

\section*{Acknowledgements}
\label{sec_acknowledgements}

The work of R.J. and G.M. was supported by Grant-in-Aid for JSPS Fellows (No.~25-8360 and No.~26-10887, respectively).
The authors are grateful to the referee for the careful reading of the manuscript and a number of excellent suggestions.

\appendix

\section{Color Factors}
\label{app_colfac}

The basic formula is
\begin{align}
(t^a)_{ij}(t^a)_{kl}
&= \genfrac{}{}{}{1}{1}{2} \left(\delta_{il}\delta_{kj}-\genfrac{}{}{}{1}{1}{N} \delta_{ij}\delta_{kl}\right),
\end{align}
where $(t^a)_{ij}$ is the generator of SU$(N)$ in fundamental representation.
In the meson channel, the ladder diagrams have a factor
\begin{align}
t^{a_1}t^{a_2}\cdots t^{a_{n-1}}t^{a_n}t^{a_n}t^{a_{n-1}}\cdots t^{a_2}t^{a_1} &= \genfrac{}{}{}{1}{1}{2^n} \left(N-\genfrac{}{}{}{1}{1}{N} \right)^n
\equiv c^{\rm{(M)}}_n,
\label{eq_meson_ladderA}
\end{align}
whereas the completely crossed ladder diagrams have 
\begin{align}
t^{a_1}t^{a_2}\cdots t^{a_n}t^{a_1}t^{a_2}\cdots t^{a_n} &= \genfrac{}{}{}{1}{1}{2^{n+1}} \left( N-\genfrac{}{}{}{1}{1}{N} \right) \left[ \left(-1-\genfrac{}{}{}{1}{1}{N} \right) ^{n-1}+\left( 1-\genfrac{}{}{}{1}{1}{N} \right) ^{n-1}\right]
\equiv d^{\rm{(M)}}_n.
\label{eq_meson_pladderA}
\end{align}
The ladder diagrams in the diquark channel have a factor
\begin{align}
t^{a_1}t^{a_2}\cdots t^{a_{n-1}}t^{a_n}\epsilon t^{Ta_n}t^{Ta_{n-1}}\cdots t^{Ta_2}t^{Ta_1} &= \genfrac{}{}{}{1}{1}{2^n} \left(-1-\genfrac{}{}{}{1}{1}{N} \right)^n\epsilon
\equiv c^{\rm{(D)}}_n \epsilon,
\label{eq_diquark_ladderA}
\end{align}
whereas the completely crossed ladder diagrams have
\begin{align}
t^{a_1}t^{a_2}\cdots t^{a_n}\epsilon t^{Ta_1}t^{Ta_2}\cdots t^{Ta_n} &= \genfrac{}{}{}{1}{1}{2^n}  \left[ (-1)^n (N+1) \genfrac{}{}{}{1}{1}{N^{n+1}} -\genfrac{}{}{}{1}{1}{N} \left( N-\genfrac{}{}{}{1}{1}{N} \right) ^{n} \right]\epsilon
\equiv d^{\rm{(D)}}_n \epsilon.
\label{eq_diquark_pladderA}
\end{align}

\section{Practical expressions of the DS and BS eqs. in the coupling expansion}
\label{app_qDSE}

Here we present concrete expression of equations used in \secref{sec_BSE}.
The DS eq. with the quark self-energy \eqref{eq_selfenergy}
becomes a pair of self-consistency equations,
\begin{align}
\frac{\mA}{s\mA^2+\mB^2}=1-\sum _{n} (-G)^n  \left(\genfrac{}{}{}{1}{3}{4}\right)^n  d_n^{\rm{(M)}} 
\sum_{j=1}^n \mathfrak{a}_j^{(n)} \mA (\mA^2 s)^{j-1} (\mB^2)^{n-j},\\
\frac{\mB}{s\mA^2+\mB^2}=m_b-\sum _{n}  (-G)^n  \left(\genfrac{}{}{}{1}{3}{4}\right)^n  d_n^{\rm{(M)}} 
\sum_{j=1}^n \mathfrak{b}_j^{(n)} \mB (\mA^2 s)^{n-j} (\mB^2)^{j-1},
\label{eq_AppAB}
\end{align}
where $s=p^2$ with $\mA(s)$ and $\mB(s)$, and the coefficients $\mathfrak{a}_j^{(n)},\mathfrak{b}_j^{(n)}$
are summarized in \tblref{tbl_DScoef}.
The solutions of these equations are plotted in \figref{fig_DSE} for the cases of up to ${\mc O}(G)$, ${\mc O}(G^2)$ and ${\mc O}(G^6)$.

\begin{table}[tbp]
\caption{\small Coefficients $\mathfrak{a}_j^{(n)}$ and $\mathfrak{b}_j^{(n)}$.}
\label{tbl_DScoef}
\begin{center}
\begin{tabular}{c|l||cccccc}
\multicolumn{2}{c||}{\Gcenter{2}{$\mathfrak{a}_j^{(n)}$}} & \multicolumn{6}{c}{$j$} \\\cline{3-8}
\multicolumn{2}{c||}{} & 1 & 2 & 3 & 4 & 5 & 6 \\\hline \hline
\multirow{6}{*}{$n$}
& \multirow{1}{*}{1} & 2 & & & & &\\
& \multirow{1}{*}{2} & -8 & -12 & & & &\\
& \multirow{1}{*}{3} & 32 & 64 & 16 & & &\\
& \multirow{1}{*}{4} & -128 & -336 & -256 & -112 & &\\
& \multirow{1}{*}{5} & 512 & 1536 & 1728 & 1152 & 192 &\\
& \multirow{1}{*}{6} & -2048 & -7360 & -12288 & -12672 & -6144 & -1472 \\
\end{tabular}
\end{center}
\begin{center}
\begin{tabular}{c|l||cccccc}
\multicolumn{2}{c||}{\Gcenter{2}{$\mathfrak{b}_j^{(n)}$}} & \multicolumn{6}{c}{$j$} \\\cline{3-8}
\multicolumn{2}{c||}{} & 1 & 2 & 3 & 4 & 5 & 6 \\\hline \hline
\multirow{6}{*}{$n$}
& \multirow{1}{*}{1} & 4 & & & & &\\
& \multirow{1}{*}{2} & -12 & -8 & & & &\\
& \multirow{1}{*}{3} & 32 & 32 & 16 & & &\\
& \multirow{1}{*}{4} & -112 & -256 & -336 & -128 & &\\
& \multirow{1}{*}{5} & 384 & 1152 & 1728 & 768 & 64 &\\
& \multirow{1}{*}{6} & -1472 & -6144 & -12672 & -12288 & -7360 & -2048 \\
\end{tabular}
\end{center}
\end{table}

In \secref{sec_BSE}, we see that 
the problem of whether a homogenous BS eq. has nontrivial solutions or not reduces to
a problem of whether a determinant equation \eqref{eq_det} has solutions or not.
The matrices we used in that section are expressed as
\begin{align}
H(s)=\sum _n (-G)^n  \left(\genfrac{}{}{}{1}{3}{4}\right)^n  d_n^{({\rm M})} h_n ^{({\rm M})} (s)
\end{align}
in the meson case, and
\begin{align}
H(s)=\sum _n (-G)^n  \left(\genfrac{}{}{}{1}{3}{4}\right)^n  d_n^{({\rm D})} h_n ^{({\rm D})} (s)
\end{align}
in the diquark case.
Here
\begin{align}
h_n^{({\rm M})}
&= 
\left(
\begin{matrix}
h_{n,(1,1)}^{({\rm M})} & h_{n,(1,2)}^{({\rm M})} \\
h_{n,(2,1)}^{({\rm M})} & h_{n,(2,2)}^{({\rm M})} \\
\end{matrix}
\right)
= 
\sum_{k=0}^{n}
\left(
\begin{matrix}
h_{n,(1,1),k}^{({\rm M})} & h_{n,(1,2),k}^{({\rm M})}(\mB/\mA) \\
h_{n,(2,1),k}^{({\rm M})}(\mA/\mB) & h_{n,(2,2),k}^{({\rm M})} \\
\end{matrix}
\right)
(\mA^2 s)^{n-k} (\mB^2)^k,
\label{eq_detMexplicit}
\end{align}
and
\begin{align}
h_n^{({\rm D})}
&= 
\left(
\begin{matrix}
h_{n,(1,1)}^{({\rm D})} & h_{n,(1,2)}^{({\rm D})} \\
h_{n,(2,1)}^{({\rm D})} & h_{n,(2,2)}^{({\rm D})} \\
\end{matrix}
\right)
= 
\sum_{k=0}^{n}
\left(
\begin{matrix}
h_{n,(1,1),k}^{({\rm D})} & h_{n,(1,2),k}^{({\rm D})}(\mB/\mA) \\
h_{n,(2,1),k}^{({\rm D})}(\mA/\mB) & h_{n,(2,2),k}^{({\rm D})} \\
\end{matrix}
\right)
(\mA^2 s)^{n-k} (\mB^2)^k,
\label{eq_detDexplicit}
\end{align}
where $s = P^2$, $\mA = \mA(s/4)$ and $\mB = \mB(s/4)$. 
The coefficients $h_{n,(i,j),k}^{({\rm M})}$ and $h_{n,(i,j),k}^{({\rm D})}$ are summarized in \tblref{tbl_BScoef}.
The values of det$(\bm{I}-H)$ calculated using the above matrices are plotted in \figref{fig_diquarktot}.

\begin{landscape}
\begin{table}[tbp]
\caption{\small Coefficients in Eqs.~\eqref{eq_detMexplicit} -- \eqref{eq_detDexplicit}.}
\label{tbl_BScoef}
\begin{minipage}{0.5\columnwidth}
\begin{center}
\scalebox{0.9}[0.9]{
\begin{tabular}{c|c||ccccccc}
\multicolumn{2}{c||}{\Gcenter{2}{$h_{n,(i,j),k}^{\rm{(M)}}$}} & \multicolumn{7}{c}{$k$} \\\cline{3-9}
\multicolumn{2}{c||}{} & 0 & 1 & 2 & 3 & 4 & 5 & 6 \\\hline \hline
$n$ & $(i,j)$ & & & & & & &\\\cline{1-2}
\multirow{4}{*}{1}
& \multirow{1}{*}{$(1,1)$} & 1 & -4 & & & & &\\
& \multirow{1}{*}{$(1,2)$} & 4 & 0 & & & & &\\
& \multirow{1}{*}{$(2,1)$} & 0 & 2 & & & & &\\
& \multirow{1}{*}{$(2,2)$} & $-1/2$ & 2 & & & & &\\\hline
\multirow{4}{*}{2}
& \multirow{1}{*}{$(1,1)$} & $-1/4$ & 3 & 8 & & & &\\
& \multirow{1}{*}{$(1,2)$} & -2 & -4 & 0 & & & &\\
& \multirow{1}{*}{$(2,1)$} & 0 & -1 & -8 & & & &\\
& \multirow{1}{*}{$(2,2)$} & $1/2$ & 3 & -4 & & & &\\\hline
\multirow{4}{*}{3}
& \multirow{1}{*}{$(1,1)$} & 1 & 6 & 12 & -16 & & &\\
& \multirow{1}{*}{$(1,2)$} & 3 & 20 & 48 & 0 & & &\\
& \multirow{1}{*}{$(2,1)$} & 0 & 3 & 28 & 48 & & &\\
& \multirow{1}{*}{$(2,2)$} & $-1/2$ & 0 & 36 & 80 & & &\\\hline
\multirow{4}{*}{4}
& \multirow{1}{*}{$(1,1)$} & $-13/16$ & $-23/4$ & -19 & 28 & 128 & &\\
& \multirow{1}{*}{$(1,2)$} & $-5/2$ & -23 & -112 & -176 & 0 & &\\
& \multirow{1}{*}{$(2,1)$} & 0 & $-11/4$ & -28 & -92 & -160 & &\\
& \multirow{1}{*}{$(2,2)$} & $1/2$ & $7/4$ & -19 & -92 & -208 & &\\\hline
\multirow{4}{*}{5}
& \multirow{1}{*}{$(1,1)$} & 1 & $33/2$ & 99 & 204 & 144 & -64 &\\
& \multirow{1}{*}{$(1,2)$} & $11/4$ & 45 & 288 & 720 & 704 & 0 &\\
& \multirow{1}{*}{$(2,1)$} & 0 & $13/4$ & 51 & 288 & 816 & 832 &\\
& \multirow{1}{*}{$(2,2)$} & $-1/2$ & $-3/2$ & 51 & 396 & 1296 & 1472 &\\\hline
\multirow{4}{*}{6}
& \multirow{1}{*}{$(1,1)$} & $-61/64$ & $-159/16$ & $-93/2$ & -78 & -36 & 720 & 2048\\
& \multirow{1}{*}{$(1,2)$} & $-21/8$ & $-139/4$ & -240 & -936 & -2528 & -3264 & 0\\
& \multirow{1}{*}{$(2,1)$} & 0 & $-51/16$ & $-79/2$ & -234 & -960 & -2224 & -2688\\
& \multirow{1}{*}{$(2,2)$} & $1/2$ & $45/16$ & $-9/4$ & -78 & -744 & -2544 & -3904\\
\end{tabular}}
\end{center}
\end{minipage}
\begin{minipage}{0.5\columnwidth}
\begin{center}
\scalebox{0.9}[0.9]{
\begin{tabular}{c|c||ccccccc}
\multicolumn{2}{c||}{\Gcenter{2}{$h_{n,(i,j),k}^{\rm{(D)}}$}} & \multicolumn{7}{c}{$k$} \\\cline{3-9}
\multicolumn{2}{c||}{} & 0 & 1 & 2 & 3 & 4 & 5 & 6 \\\hline \hline
$n$ & $(i,j)$ & & & & & & &\\\cline{1-2}
\multirow{4}{*}{1}
& \multirow{1}{*}{$(1,1)$} & -1 & 4 & & & & &\\
& \multirow{1}{*}{$(1,2)$} & -4 & 0 & & & & &\\
& \multirow{1}{*}{$(2,1)$} & 0 & -2 & & & & &\\
& \multirow{1}{*}{$(2,2)$} & $1/2$ & -2 & & & & &\\\hline
\multirow{4}{*}{2}
& \multirow{1}{*}{$(1,1)$} & $-1/4$ & 3 & 8 & & & &\\
& \multirow{1}{*}{$(1,2)$} & -2 & -4 & 0 & & & &\\
& \multirow{1}{*}{$(2,1)$} & 0 & -1 & -8 & & & &\\
& \multirow{1}{*}{$(2,2)$} & $1/2$ & 3 & -4 & & & &\\\hline
\multirow{4}{*}{3}
& \multirow{1}{*}{$(1,1)$} & $-1/4$ & 0 & 0 & 16 & & &\\
& \multirow{1}{*}{$(1,2)$} & $-3/2$ & -8 & -24 & 0 & & &\\
& \multirow{1}{*}{$(2,1)$} & 0 & $-3/2$ & -16 & -24 & & &\\
& \multirow{1}{*}{$(2,2)$} & $1/2$ & 3 & -12 & -32 & & &\\\hline
\multirow{4}{*}{4}
& \multirow{1}{*}{$(1,1)$} & $-1/16$ & $1/4$ & 5 & 76 & 128 & &\\
& \multirow{1}{*}{$(1,2)$} & -1 & -11 & -64 & -80 & 0 & &\\
& \multirow{1}{*}{$(2,1)$} & 0 & $-5/4$ & -16 & -44 & -64 & &\\
& \multirow{1}{*}{$(2,2)$} & $1/2$ & $19/4$ & 5 & 4 & -16 & &\\\hline
\multirow{4}{*}{5}
& \multirow{1}{*}{$(1,1)$} & $-1/16$ & $-3/2$ & -9 & 36 & 96 & 64 &\\
& \multirow{1}{*}{$(1,2)$} & $-7/8$ & -15 & -108 & -240 & -224 & 0 &\\
& \multirow{1}{*}{$(2,1)$} & 0 &$-11/8$ & -21 & -108 & -336 & -352 &\\
& \multirow{1}{*}{$(2,2)$} & $1/2$ & $21/4$ & 9 & -36 & -336 & -512 &\\\hline
\multirow{4}{*}{6}
& \multirow{1}{*}{$(1,1)$} & $-1/64$ & $-27/16$ & $-21/2$ & 66 & 492 & 1680 & 2048\\
& \multirow{1}{*}{$(1,2)$} & $-3/4$ & $-73/4$ & -168 & -648 & -1472 & -1344 & 0\\
& \multirow{1}{*}{$(2,1)$} & 0 & $-21/16$ & -23 & -162 & -672 & -1168 & -768\\
& \multirow{1}{*}{$(2,2)$} & $1/2$ & $105/16$ & $123/4$ & 66 & -168 & -432 & -64\\
\end{tabular}}
\end{center}
\end{minipage}
\end{table}
\end{landscape}

\section{Comment on the number of the independent elements of the basis}
\label{app_proof}

The following four elements made of $\slashed{P}$ and $\gamma$'s in Eq.~\eqref{eq_basis}
\begin{align}
&\gamma_\mu\gamma_\nu\gamma_\rho\gamma_\sigma \otimes \gamma_\mu\gamma_\nu\gamma_\rho\gamma_\sigma, \qquad \quad
\gamma_\mu\gamma_\nu\gamma_\rho\gamma_\sigma \otimes \slashed{P}\gamma_\mu\gamma_\nu\gamma_\rho\gamma_\sigma, \nn\\
&\slashed{P}\gamma_\mu\gamma_\nu\gamma_\rho\gamma_\sigma \otimes \gamma_\mu\gamma_\nu\gamma_\rho\gamma_\sigma, \qquad
\slashed{P}\gamma_\mu\gamma_\nu\gamma_\rho\gamma_\sigma \otimes \slashed{P}\gamma_\mu\gamma_\nu\gamma_\rho\gamma_\sigma, \nn
\end{align}
may seem independent of the other 16 elements in Eq.~\eqref{eq_basis}, but these are not since the following equation holds:
\begin{align}
\gamma_\mu \gamma_\nu \gamma_\rho \gamma_\sigma \otimes \gamma_\mu \gamma_\nu \gamma_\rho \gamma_\sigma
 =& -\frac{16}{P^2}\slashed{P}\gamma_\mu \otimes \slashed{P}\gamma_\mu
+ 4\gamma_\mu \gamma_\nu \otimes \gamma_\mu \gamma_\nu 
+ \frac{4}{P^2} \slashed{P} \gamma_\mu \gamma_\nu \gamma_\rho \otimes \slashed{P} \gamma_\mu \gamma_\nu \gamma_\rho .
\label{eq_basisrel}
\end{align}
To check this, note that Eq.~(\ref{eq_basisrel}) is rewritten as
\begin{align}
 \left( \delta^{\alpha \beta}  
 - 4P^\alpha P^\beta /P^2 \right) 
\left( \gamma_\alpha \gamma_\mu\gamma_\nu\gamma_\rho \otimes \gamma_\beta \gamma_\mu\gamma_\nu\gamma_\rho 
- 4\gamma_\alpha\gamma_\mu \otimes \gamma_\beta\gamma_\mu \right)   = 0.
\label{eq_basisrel2}
\end{align}
This is equivalent to 
\begin{align}
\left( \delta^{\alpha \beta} 
- 4P^\alpha P^\beta /P^2 \right) 
\left( \gamma_\alpha \gamma_\mu\gamma_\nu\gamma_\rho M \gamma_\beta \gamma_\mu\gamma_\nu\gamma_\rho 
- 4\gamma_\alpha\gamma_\mu M \gamma_\beta\gamma_\mu \right) = 0.
\label{eq_basisrec3}
\end{align}
for an arbitrary Hermitian $4 \times 4$ matrix $M$, 
which can be expanded by $\gamma_{\mu_1}...\gamma_{\mu_n}$. 
One can readily check that the second parentheses of Eq.~(\ref{eq_basisrec3}) vanishes when $n$ is odd, 
while it is proportional to $\delta_{\alpha \beta}$ 
when $n$ is even and vanishes after it is multiplied by the first parenthesis.



\end{document}